\documentstyle[12pt]{article}
\begin{document}
\begin{center}
\vfill
\large\bf{Vortex Solutions in the\\
Chern-Simons Stuekelberg Model}\\
\end{center}
\vfill
\begin{center}
D.G.C. McKeon\\
Department of Applied Mathematics\\
University of Western Ontario\\
London ~CANADA\\
N6A 5B7\\
\vfill
\end{center}
\vfill
email:  TMLEAFS@APMATHS.UWO.CA\\
Fax: 519-661-3523\\
Telephone: 519-679-2111, Ext. 8789 \hfill PACS no.: 11.15Kc
\eject
\section{Abstract}
Vortex solutions to the classical field equations in a massive, 
renormalizable $U(1)$ gauge model
are considered in (2+1) dimensions.  A vector field whose kinetic 
term consists of a
Chern-Simons term plus a Stuekelberg mass term 
is coupled to a scalar field.  If the classical scalar
field is set equal to zero, then there are classical configurations 
of the vector field in which the
magnetic flux is non-vanishing and finite.  In contrast to the 
Nielsen-Olesen vortex, the magnetic
field vanishes exponentially at large distances and diverges 
logarithmicly at short distances.  This
divergence, although not so severe as to cause the flux to diverge, 
results in the
Hamiltonian
becoming infinite.  If the classical scalar field is no longer equal 
to zero, then the magnetic flux
is not only finite, but quantized and the asymptotic behaviour of the 
field is altered so that the
Hamiltonian no longer suffers from a divergence due to the field 
configuration at the origin. 
Furthermore, the asymptotic behaviour at infinity is dependent on the 
magnitude of the
Stuekelberg mass term.

\section{Introduction}
Static vortex solutions of the $3 + 1$ dimensional Abelian Higgs 
model (1) have been known
for some time.  The $2 + 1$ dimensional vortex solutions in models 
with a pure Chern-Simons
kinetic term for the $U(1)$ gauge field have also been discussed (2-
4) (see also (5)).  This
motivates us to consider vortex solutions in a $2 + 1$ dimensional 
model in which there is both
a Chern-Simons and Stuekelberg mass term for the $U(1)$ gauge field.  
This theory has been
shown to be a renormalizable model for a massive vector field (6).  
The matter field to which
this vector field couples is taken to be a scalar with quadratic and 
quartic self-interactions.

We find that even when the scalar field vanishes, this model supports 
a vortex-like solution. This
configuration has the interesting property that at infinity, the 
vector field $A_\mu (\vec{r})$ dies
out exponentially fast provided the Stuekelberg mass is non-zero.  
Consequently the contour
integral ${\displaystyle{\oint}} \, \vec{A} (\vec{r}) \cdot d 
\vec{\ell}$ about a circle at spatial
infinity is zero,
even though the flux $\Phi = {\displaystyle{\int}} \; dS \; 
\epsilon_{ij} \partial_i A_j$ is finite
and non-zero.  This
is due to a logarithmic singularity in $\epsilon_{ij} \partial_i A_j$ 
at the origin, which precludes
equating these two integrals using Stokes' theorem.  This same 
singularity at the origin results
in the Hamiltonian density behaving like $r^{-2}$ as $r$ approaches 
zero.

In order to excise this divergence, we allow the scalar field to be 
non-vanishing.  The flux now
is quantized and the Hamiltonian free of divergences due to the 
singular behaviour at the origin,
although the vector field exhibits logarithmic behaviour near $r = 
0$.  Finiteness at infinity is
maintained provided the asymptotic behaviour is chosen appropriately, 
depending on the
magnitude of the Stuekelberg mass.

\section{The Model}

As in (6), we consider the action
$$S = \int\; d^3 x \left\lbrace \frac{1}{2} \,\epsilon^{\mu\alpha\nu} 
A_\mu \partial_\alpha
A_\nu -
 \frac{1}{2}\, \mu (A_\mu + \partial_\mu \sigma)^2 \right. \nonumber$$
$$\left. + \frac{1}{2} | (\partial_\mu + ieA_\mu)\phi|^2 + c_2
\phi^*\phi - c_4(\phi^*
\phi)^2 \right\rbrace\eqno[1]$$
$$(g_{\mu\nu} = (+\;\;+\;\;-);\;\;\epsilon_{012} = +1;\;\;\; (\mu, 
c_2, c_4) > 0).\nonumber$$
In the first instance, we set $\phi = 0$ and make the ansatz
$$A_\mu (\vec{r}) = \left( - 
\frac{A(r)y}{r^2}\;,\;\;\frac{A(r)x}{r^2}\;,\;\;A_0(r) 
\right)\eqno[2]$$
in the gauge in which the Stuekelberg field $\sigma$ vanishes.  The 
field equations
$$\epsilon_{ij} \partial_i A_j = \mu A_0\eqno[3a]$$ 
$$\;\;\;\epsilon_{ij} \partial_i A_0 = -\mu A_i\eqno[3b]$$
then reduce to
$$A^\prime (r) = \mu r A_0(r)\eqno[4a]$$
$$A_0^\prime (r) = \frac{\mu A(r)}{r}\;.\eqno[4b]$$
>From [4] it is easily seen that
$$A^{\prime\prime}_0 + \frac{1}{r}\; A_0^\prime - \mu^2 \; A_0 = 0 
\eqno[5]$$
whose solution is given in terms of associated Bessel functions
$$A_0(r) = \alpha K_0 (\mu r) + \beta I_0 (\mu r)\;\; . \eqno[6]$$
Setting $\beta = 0$ in order to ensure that $A_0(r)$ vanishes at 
infinity, we see from [6] and [4b]
that
$$A(r) = -\alpha r \; K_1 (\mu r)\;\; . \eqno[7]$$

(The integral representation
$$K_\nu(x) = \int_0^\infty \; dt\; e^{-x \cosh t} \cosh \nu 
t\eqno[8]$$
for $K_\nu (x)$ is useful.)  The constant of integration $\alpha$ in 
[6] and [7] is not fixed by
considerations related to the equations [4].  (In (2-4), the constant 
of integration analogous to
$\alpha$ is determined by requiring finiteness at $r = 0$; this is 
not possible here.) 

The total flux of magnetic field through the two-dimensional space of 
our model is
$$\Phi = \int \; d^2r \; \epsilon_{ij} \partial_i A_j\eqno[9]$$
which by [2], [3a] and [6] becomes
$$\hspace{2.3cm}= \mu \alpha \int_0^{2\pi} \, d\theta \,\int_0^\infty
\, dr\;r \;\; K_0(\mu r)\nonumber$$
$$\hspace{-1cm}= \frac{4\alpha \pi}{\mu}\;\; . \eqno[10]$$
There consequently is a non-vanishing magnetic flux in this model 
which is finite provided $\mu
\neq 0$.  However, Stoke's theorem cannot be used to rewrite [9] in 
terms of a line integral
$$\tilde{\Phi} = \oint_c \, d \vec{\ell} \cdot \vec{A} (\vec{r}) 
\eqno[11]$$
where the contour $C$ in [11] is a circle at infinity, as $\vec{A}
(\vec{r})$ is singular at $r = 0$,
as follows from [2] and [7].  Indeed, since $K_\nu(x) \rightarrow 
\sqrt{\frac{\pi}{2x}} e^{-x}$
as $x \rightarrow \infty$, we see that $\tilde{\Phi} = 0$ in [11].

In (6), it is shown that the Hamiltonian density associated with 
action of [1] is given by
$${\cal{H}} = - \frac{\mu}{2} \; A_0^2 + \frac{1}{2\mu} 
\left(\pi_\sigma + \frac{\mu}{2}
A_0\right)^2 + A_0
\epsilon_{ij} \partial_i A_j + \frac{\mu}{2} \left( \partial_i \sigma 
+
A_i\right)^2\nonumber$$
$$+ \frac{1}{2} |\pi_\phi|^2 + \frac{1}{2} \left| \left( \partial_i + 
ie
A_i \right)\phi\right|^2 - c_2 \phi^*\phi +
c_4(\phi^*\phi)^2\eqno[12]$$
once the scalar field is treated in the standard fashion.  (Here we 
have $\pi_\phi =
(\partial_0 + ieA_0)\phi$ and $\pi_\sigma = -\mu(\partial_0 \sigma + 
\frac{1}{2} A_0)$.)  Taking
into account the ansatz of [2] and the equation of motion of [3a], we 
see that [12] reduces to
$${\cal{H}} = \frac{\mu}{2} \left(A_0^2 + \vec{A}^2\right)\eqno[13]$$
which, in turn becomes
$$\hspace{2cm}= \frac{\mu\alpha^2}{2} \left[ K_0^2 (\mu r) + K_1^2 
(\mu r)\right]\eqno[14]$$
as follows from [2], [6] and [7].  Since $K_0(x) \sim -\log x$ and 
$K_1(x) \sim \frac{1}{x}$
near $x = 0$, we see that the energy
$$E = \int \,d^2r {\cal{H}} \eqno[15]$$
diverges for this field configuration.

We now examine classical field configurations in which the scalar 
field $\phi$ in [1] is non-zero. 
Supplementing the ansatz of [2] with
$$\phi (r,\theta) = e^{in \theta} f(r)\;\;\;(n = 0, 1 \ldots, \theta 
= \tan^{-1}
\frac{y}{x})\eqno[16]$$
then the linear equations of [4] becomes
$$-e(n + eA)f^2 = r A_0^\prime - \mu\left(A + 
\frac{n}{e}\right)\eqno[17a]$$
$$-e^2 A_0f^2 = \frac{1}{r} A^\prime - \mu A_0\eqno[17b]$$
$$f^{\prime\prime} + \frac{1}{r} f^\prime + e^2 A_0^2 f - \frac{(n +
eA)^2f}{r^2} + 2c_2 f -
4c_4f^3 = 0\eqno[17c]$$
provided the Stuekelberg field is taken to be
$$\sigma = \frac{n}{e}\, \theta\; .\eqno[18]$$
The boundary conditions to these equations are taken to be
$$\lim_{r \rightarrow \infty} (A(r),\, A_0(r),\, f(r)) = \left( -
\frac{n}{e},\, 0,\, \sqrt{c_2/4c_4} \right)
\eqno[19]$$
and
$$\lim_{r \rightarrow 0} (A(r), A_0(r), f(r)) = \left( \lambda 
r^\alpha \ln
r,\, \lambda_0
r^{\alpha_{0}}\ln r,\, \sigma r^\beta \right). \eqno[20]$$
The logarithmic terms in [20] are normally not encountered in vortex 
solutions (1-5), but do not
affect the finiteness of either the magnetic flux or energy of the 
field configuration.  The
asymptotic behaviour that we postulate is, for small $r$
$$(A(r), A_0(r), f(r)) \sim (\lambda r^2 \ln r, \,\lambda_0 \ln r,\, 
\sigma r^\beta).\eqno[21]$$
Substitution of [21] into the field equations [17] and examining 
leading order terms at the origin
shows that
$$\lambda_0 = \frac{\mu n}{e} \eqno[22a]$$
$$\lambda = \frac{\mu^2 n}{2e}\eqno[22b]$$
$$\beta = |n|\; .\eqno[22c]$$
Similarly, as $r \rightarrow \infty$, the field equations [17] are 
satisfied to leading order provided
either
$$(A(r), A_0(r), f(r)) \sim \left( - \frac{n}{e} + a_1
e^{-e^2\left(\frac{c_2}{2c_4} -
\frac{\mu}{e^2}\right)r},\, \frac{a_1}{r} e^{-
e^2\left(\frac{c_2}{2c_4} -
\frac{\mu}{e^2}\right)r},\, \sqrt{\frac{c^2}{2c_4}} + \ldots a_2 e^{-
\sqrt{4c_2}r} \right)\nonumber$$
$$\left( {\rm{if}} \; \frac{c_2}{2c_4} > \frac{\mu}{e^2} 
\right)\eqno[23a]$$
or
$$(A(r), A_0(r), f(r)) \sim \left( - \frac{n}{e} + a_1 e^{-
e^2\left(\frac{\mu}{e^2} -
\frac{c_2}{2c_4} \right)r},\, -\frac{a_1}{r} e^{-
e^2\left(\frac{\mu}{e^2} -
\frac{c_2}{2c_4} 
\right)r},\, a_2 e^{-\sqrt{4c_2}r} \right)\nonumber$$
$$\left( {\rm{if}} \; \frac{c_2}{2c_4} < \frac{\mu}{e^2} 
\right).\eqno[23b]$$
The asymptotic behaviour at infinity is consequently dependent on the 
magnitude of $\mu$.  In
either case though, the magnetic flux through space is given by
$$\Phi = \int \, d^2x \; \epsilon_{ij} \partial_{i} A_j \nonumber$$
$$\hspace{.5cm} = - \int\; d^2x \, \frac{1}{r} \, 
\frac{dA}{dr}\nonumber$$
$$\hspace{-.7cm} = \frac{2\pi n}{e}\; .\eqno[24]$$
Consequently the magnetic flux quantized in units of 
$\displaystyle{\frac{2\pi}{e}}$, as in (1-4).

\section{Discussion}
We have considered static, vortex-like solutions in a $2 + 1$ 
dimensional $U(1)$ gauge theory
in which both a Chern-Simons and Stuekelberg term occurs.  An 
explicit solution with finite flux
and divergent energy can be constructed when there is no scalar 
field.  In the presence of a 
non-vanishing scalar field, the asymptotic form of solutions to the 
field equations can be found
if the
Stuekelberg mass is non-zero; it differs qualitatively from the form 
of solutions occurring when
this mass vanishes.  Nevertheless, these solutions have finite energy 
and quantized flux.
Computing the charge and angular momentum of these solutions would 
not be straight forward.

It is of interest to see if these solutions have practical 
application in condensed matter physics.

\section{Acknowledgement}
We would like to thank F. Dilkes for discussion and NSERC for 
financial support.


\begin{thebibliography}{99}
\bibitem{1} H.B. Nielsen and P. Olesen, Nucl. Phys. B61 (1973) 45 
(reprinted in: ``Solitons and
Particles'' eds. C. Rebbi and G. Soliani, World Scientific, Singapore 
1984).
\bibitem{2} D. Jatkar and A. Khare, Phys. Lett. B236 (1990) 283.
\bibitem{3} J. Hong, V. Kim and P.Y. Pac, Phys. Rev. Lett. 64 (1990) 
2230.
\bibitem{4} R. Jackiw and E. Weinberg, Phys. Rev. Lett. 64 (1990) 
2234.
\bibitem{5} A. Antill\'{o}n, J. Escabna, and M. Torres, Phys. Rev. 
D55 (1997) 6327.
\bibitem{6} F.A. Dilkes and D.G.C. McKeon, Phys. Rev. D52 (1995) 4668.
\end{thebibliography}
\end{document}